\documentclass[twocolumn,superscriptaddress]{revtex4}
\usepackage[dvips]{graphicx}
\usepackage{amssymb,amsfonts,amsmath}
\usepackage{color}
\usepackage{eurosym}
\usepackage[sort&compress]{natbib}

\begin{document}

\title{\sffamily Human strategy updating in evolutionary games}

\author{Arne Traulsen}
\email{traulsen@evolbio.mpg.de}
\address{Emmy-Noether Group for Evolutionary Dynamics, Max Planck Institute for Evolutionary Biology, 24306 Pl{\"o}n, Germany}
\address{Department of Evolutionary Ecology, Max Planck Institute for Evolutionary Biology, 24306 Pl{\"o}n, Germany}
\author{Dirk Semmann}
\address{Department of Evolutionary Ecology, Max Planck Institute for Evolutionary Biology, 24306 Pl{\"o}n, Germany}
\address{Junior Research Group Evolution of
Cooperation and Prosocial Behavior, Courant Research Center Evolution of Social Behavior, 37077 G{\"o}ttingen, Germany}
\author{Ralf D. Sommerfeld}
\address{Department of Evolutionary Ecology, Max Planck Institute for Evolutionary Biology, 24306 Pl{\"o}n, Germany}
\author{Hans-J{\"u}rgen Krambeck}
\address{Department of Evolutionary Ecology, Max Planck Institute for Evolutionary Biology, 24306 Pl{\"o}n, Germany}
\author{Manfred Milinski}
\address{Department of Evolutionary Ecology, Max Planck Institute for Evolutionary Biology, 24306 Pl{\"o}n, Germany}

\sffamily

\begin{abstract}
Evolutionary game dynamics describes not only frequency dependent genetical evolution, but also cultural evolution in humans. 
In this context, successful strategies spread by imitation.  
It has been shown that the details of strategy update rules can have a crucial impact on evolutionary dynamics in theoretical models
and e.g. significantly alter the level of cooperation in social dilemmas.
But what kind of strategy update rules can describe imitation dynamics in humans?
Here, we present a way to measure such strategy update rules in a behavioral experiment. 
We use a setting in which individuals are virtually arranged on a spatial lattice. 
This produces a large number of different strategic situations from which we can
assess strategy updating. 
Most importantly, spontaneous strategy changes corresponding to mutations or exploration behavior are more frequent than
assumed in many models. 
Our experimental approach to measure properties of the update mechanisms used in theoretical models 
will be useful for mathematical models of cultural evolution.
\end{abstract}

\maketitle


Classical game theory assumes that agents make rational decisions, taking into account that they are interdependent with other agents that are also fully rational \cite{neumann:1944ef}. While this assumption has proven to be problematic even in humans, evolutionary game theory has been developed to describe the dynamics of genetical or cultural evolution when fitness is not fixed, but depends on the interactions with others. 
Applications of this framework range from the dynamics
of microbes \cite{turner:1999hp,kerr:2002xg,rainey:2003an} to animal behavior \cite{maynard-smith:1973to,bshary:2006aa} and  human behavior \cite{bendor:1995aa,milinski:2006aa,pfeiffer:2006pf}. 
Many aspects of evolutionary dynamics hinge upon the microscopic rules describing how successful strategies spread. 
In particular in structured populations, these rules can crucially alter the evolutionary outcome
and, for example, determine whether cooperation evolves or not
 \cite{ohtsuki:2006na,szabo:2007aa,roca:2009aa}.
Thus, it is of great importance to infer how strategies are actually adopted. 
To this end, we have developed a behavioral experiment that mimics typical properties of theoretical models, but replaces the computer agents by real human players. 
Each player interacts only with her immediate neighbors. 
To evaluate his performance, each player can compare his payoff to the payoff of the neighbors and use this as a basis to adopt new strategies. 
However, there are some subtle differences between mathematical models and human behavior:
Humans may use mixed strategies, i.e.\ randomize between their options, or even change their strategies over time,
whereas most theoretical models consider the simplest case in which a player's strategy is equated with his action. 
Thus, any change in behavior is equated to a change in the strategy.
If we aim to apply this simple framework of one-shot games as a first approximation to describe human behavior, we have to infer the details of strategy adoption, e.g.\ the rate of spontaneous strategy changes. 
We utilize a spatial game in which human players are interacting with their immediate neighbors only. 
This leads to a large number of different strategic situations that allows us to infer under which circumstances a neighboring strategy is adopted.

A large portion of the literature on evolutionary games focuses on the Prisoner's Dilemma. 
This is a paradigm to study the evolution of costly cooperation among selfish individuals, because it highlights the potential differences between individual interests and the social optimum \cite{rapoport:1965pd,dawes:1980aa,axelrod:1984yo,kollock:1998aa,macy:2002hc}. 
In the Prisoner's Dilemma, two players have to decide simultaneously whether to cooperate with the other or not. 
If both players cooperate, they obtain a reward $R$. 
If one defects while the other cooperates, the defector gets $T$ (temptation to defect) and the cooperator obtains $S$ (suckerÕs payoff).
If both defect, they get a punishment $P$. 
This can be summarized by the payoff matrix
\begin{align}\label{eq:Pmatrix}
\bordermatrix{
  & C & D \cr
C & R & S \cr
D & T & P \cr}.
\end{align}
The Prisoner's Dilemma is characterized by the payoff ranking $T > R > P > S$ 
(and in addition $2R>T+S$ for repeated games). 
In this case, rational individuals choose defection: They are greedy and try to exploit other cooperators $(T>R)$, but they also fear that the other one will try to exploit them $(P>S)$. 
However, since mutual cooperation yields a higher payoff than mutual defection $(R>P)$, players face a dilemma: Individual reasoning leads to defection, but mutual cooperation implies a higher payoff. 
Similarly, in an evolutionary setting the higher payoff of defectors implies more reproductive success and thus cooperation should not evolve. However, cooperation can evolve for example by kin selection, spatial structure or when interactions are repeated \cite{nowak:2006pw}. 
There is a large body of literature on behavioral experiments based on the repeated Prisoner's Dilemma, see e.g. \cite{kagel:1997bb,camerer:2003bo}. It is clear the humans behave in a more sophisticated way than simple computer programs \cite{kagel:1997bb}, 
but it has also been shown that working memory constraints limit human behavior in repeated games \cite{milinski:1998zp}. 
Nonetheless, with few exceptions, see e.g.\ \cite{lindgren:1994to}, theorists have focused on simple forms of strategy choice, e.g. to disentangle the effects of population structure and game characteristics.
In particular, the spatial version of the Prisoner's Dilemma has been analyzed in great detail by theorists
\cite{nowak:1992pw,hauert:2001ag,skyrms:2003aa,hauert:2004bo,helbing:2009ud}.
Initially, research has focused on simple lattices that approximate interactions in spatially homogeneous systems.
More recently, many studies 
have addressed complex social networks instead \cite{abramson:2001nx,santos:2006pn}. 
Typically, players are arranged on a social network and interact pairwise only with their immediate neighbors, choosing either cooperation or defection for all interactions. 
In each round, the payoff of every player is accumulated in pairwise encounters with all its neighbors.
Individuals with high payoffs are either imitated more often than others (in social models) or produce more offspring (in genetic models). 
The dynamics in spatially structured populations depend crucially on the details of the microscopic rules by which the players update their strategies. 
Our goal is to shed some light onto these microscopic rules that describe how players change their strategies. 

Such a behavioral experiment with humans can only be done in comparably small systems due to some restrictions in experimental games that are absent in mathematical models. 
For example, participants have to be paid in real money and their anonymity must be guaranteed such that the results are not blurred
by potential reputation effects.
Throughout this study, we focus on 
$R = 0.30$ \EUR\ ,
$S = 0.00$ \EUR\ ,
$T = 0.40$ \EUR\ , and
$P = 0.10$ \EUR\ .
This leads to the $2 \times 2$ payoff matrix 
\begin{align}\label{eq:Pmatrix}
\bordermatrix{
  & C & D \cr
C & 0.30 \hbox{\, \EUR\ } & 0.00 \hbox{\, \EUR\ } \cr
D & 0.40 \hbox{\, \EUR\ } & 0.10 \hbox{\, \EUR\ } \cr}.
\end{align}
Players were virtually arranged on spatial $4 \times 4$ lattice with periodic boundary conditions,
which corresponds to the surface of a torus.
The participants had four fixed neighbors throughout the entire game.
Thus, the possible cooperator payoffs accumulated in their four interactions are 
$0.00$ \EUR , $0.30$ \EUR, $0.60$ \EUR, $0.90$ \EUR, and $1.20$ \EUR. 
A defector has the possible payoff values 
$0.40$ \EUR, $0.70$ \EUR, $1.00$ \EUR, $1.30$ \EUR, and $1.60$ \EUR.

Many theoretical studies are based on synchronous updating,  
which means that all players make strategy revisions at the same time.
This can easily be mimicked in behavioral experiments.  
However, the way that strategies are changed is more difficult to address. 
A typical assumption is that each player chooses the strategy that obtains
the highest payoff in the neighborhood, either his previous strategy or a different one. 
In our experiment, players have many different possibilities for strategy updating. 
It is clear that human players sometimes do not follow this 
 ``imitate the best`` rule, but choose their strategies in a different fashion. 
 Nonetheless, this imitation dynamics can serve as a first approximation for 
 strategy updating.

More recent studies have stressed that strategy adoption is stochastic, which can be modeled introducing an intensity of selection \cite{nowak:2004pw,tarnita:2009jx}. 
One possibility is the following imitation process with errors: 
Each player compares his payoff to the best performing neighbor that has played a different strategy and calculates the payoff difference $\Delta \pi$.  
With probability  $p= (1+\exp[-\beta \Delta \pi ])^{-1} $, he adopts the neighbors strategy \cite{blume:1993jf,szabo:1998wv,traulsen:2007cc}. 
Here, $\beta$ measures the intensity of selection, i.e. how important the payoffs are for strategy revisions. 
In our case with two strategies only, this is equivalent to the multinomial logit model \cite{mcfadden:1981aa,sandholm:2010bo}.
Our goal is to understand which strategy adaption rules can describe human behavior in this game.

\section{Results}

\begin{figure}[h]
\begin{center}
	\includegraphics[width=\linewidth,angle=0]{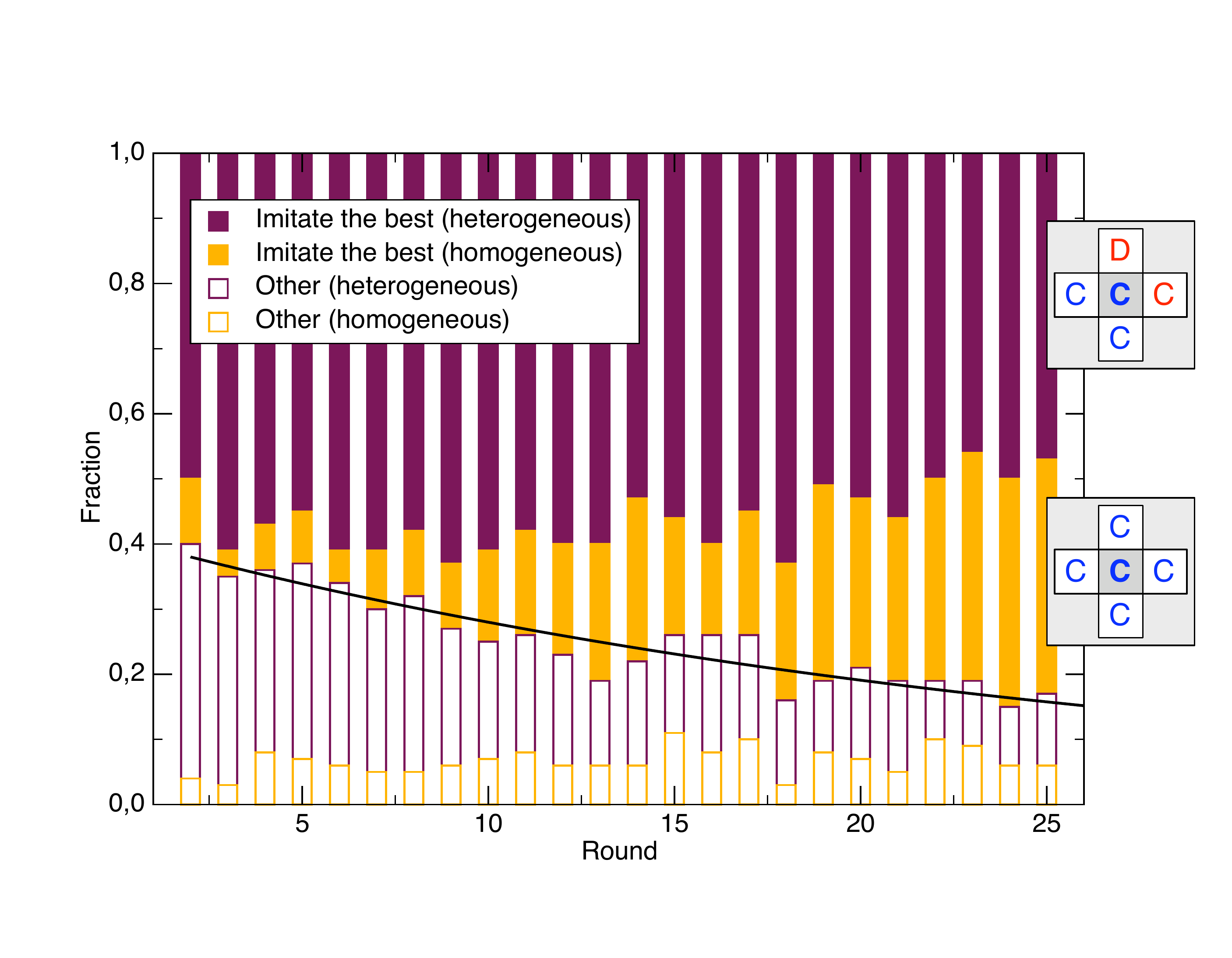}
	\end{center}
\caption{ 
Strategy updating in behavioral experiments with fixed neighbors.
Most strategy changes can be explained by imitation of the most successful
neighbor (full bars), i.e.\ changing to the best available strategy in a heterogeneous environment where neighbors
play different strategies (brown) or sticking to the strategy when everyone does the same in a homogeneous environment (orange). 
However, a large portion of strategy changes cannot be explained by imitation (open bars). 
These are either spontaneous strategy switching in homogeneous environments 
with no role model (open orange bars) or choosing 
a strategy that did not perform best in an environment with different neighboring strategies (open brown bars). 
The line shows a fit of the fraction of strategy changes not explained by imitation.
This fraction decays approximately exponentially as $\nu_{0} \cdot \Gamma^{t-1}$. 
A nonlinear regression leads to $\nu_{0} = 0.380 \pm 0.013$ and $\Gamma = 0.962 \pm 0.003$  (full line).
The diagrams on the right show an example for a heterogeneous (top) and a homogeneous environment (bottom) of a focal cooperating player. In total, we have 4315 heterogeneous  situations and 1445 homogeneous situations in our 5760 strategy choice situations
(graphic shows averages over 15 fixed neighbor treatments with 25 rounds and 16 player each).
}
\label{figtime}
\end{figure}

Let us first address if imitation dynamics can describe human strategy updating. 
In total, we have 5760 individual decisions to keep a strategy or to switch it. 
As a first model, we assume that all individuals use the imitate the best rule, i.e. they always imitate 
the best performing neighbor strategy, including their own. 
It has been shown that this cannot fully describe human behavior \cite{kirchkamp:2007aa}.
Fig. \ref{figtime} reveals that in our experiment, initially 62\% of the individuals follow the imitate the best rule. 
However, the remaining 38\% of the strategy changes cannot be explained by pure imitation. 
This fraction tends to decrease over time in the experiment. 
Fitting an exponential function to the data from Fig.~\ref{figtime} reveals that the fraction of strategy choices that are not explained by imitation decreases roughly by 4\% per round. 
This reflects the fact that strategy choice changes over time in our behavioral experiment
and that a stationary state is not reached.

\begin{figure}[h]
\begin{center}
	\includegraphics[width=\linewidth,angle=0]{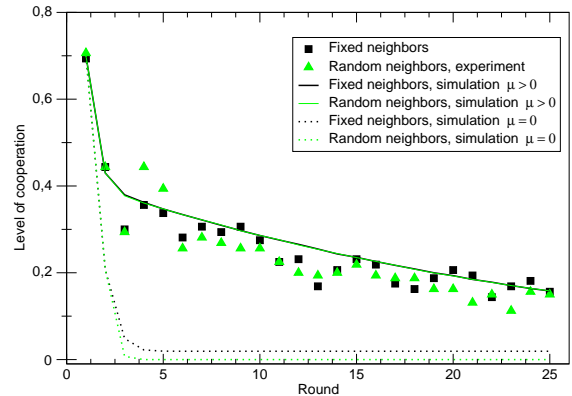}
	\end{center}
\caption{ 
The average level of cooperation tends to decrease over time. 
Symbols show a behavioral experiment with humans and lines correspond to simulations.  
In the experiment, the treatment with fixed neighbors on a $4 \times 4$ lattice with periodic boundary conditions (squares) is not significantly different from the dynamics in a system with random neighbors (triangles). 
Full lines show computer simulation in which players either imitate their best performing neighbor or choose a random strategy with probability $2 \nu \cdot \Gamma^{t-1}$, where $\nu = 0.38$ and $\Gamma = 0.96$ (fitted to the behavioral experiment, see text). 
For such high probability of random strategy choice, the simulation results for fixed and random neighbors are almost indistinguishable, the level of cooperation is driven by random strategy choice rather than by spatial structure. 
For comparison, dotted lines show computer simulations with no mutations
(experimental average over 15 repeats for fixed neighbors and 10 repeats for random neighbors, each with 16 players; simulations starting from the cooperation level of the experiment, averaged over $10^4$ realizations). 
}
\label{figspatial}
\end{figure}

In theoretical models of the spatial Prisoner's Dilemma, one is typically interested in the average level of cooperation of the system. 
The idea is that in a spatial setting, clusters of cooperators can form, leading to a significant degree of cooperation \cite{szabo:2007aa,roca:2009aa,nowak:1992pw,hauert:2001ag}. 
To explore how the level of cooperation is affected by spatial structure, we have also conducted a control experiment in which the spatial structure was broken up by reassigning each player's neighbors each round. Since in the spatial treatment individuals interact always with the same co-players and can form stable clusters of cooperators, one would expect a higher level of cooperation in the fixed-neighbors than in the random-neighbors treatment.
As described in previous human behavioral experiments \cite{milinski:2002ef} (and not necessarily in line with the expectations of theoreticians), the average level of cooperation at the start of the experiment is comparably large and very similar in the treatment with fixed neighbors (70.0\%, averaged over 15 repeats) and the treatment with random neighbors (70.6\%, averaged over 10 repeats). 
Most interestingly, we do not find a significant difference in the level of cooperation during the course of the game between the two treatments, see Fig.~\ref{figspatial}. 
Only in round 4, there is a significant difference between the levels of cooperation, which disappears after Bonferroni correction for multiple comparison. 
Stable clusters of cooperators are not found in our behavioral experiments.
The high probability of spontaneous strategy changes decreases the influence of spatial structure. 

It turns out that the dynamics can be explained based on the way that our subjects revise their strategies. 
The general dynamics of the system can be captured by a simple random strategy choice approach \cite{traulsen:2009aa}.
We assume that a player can do two things when revising her strategy
(i) with probability $\nu$, she chooses a random strategy, and
(ii) with probability $1-\nu$, she imitates her best performing neighbor. 
In our behavioral experiment, we find that $\nu$ decays exponentially with the round $t$ of the game as $\nu = \nu_{0} \Gamma^{t-1}$. 
Such an exponential decay of exploration rates has been reported before \cite{helbing:2004aa}.   
Our experiment yields for the best fit $\nu_{0} = 0.380$ and $\Gamma = 0.962$. 
To test our assumption, we simulated the temporal dynamics of 15 runs under imitation dynamics with four neighbors, fitting the
strategy choice parameters to the experiment. 
In order to be consistent with random strategy choice, we assume that only a fraction of $1-2\nu$ is correct imitation. 
A fraction $\nu$ is random strategy choice leading to the ``correct'' strategy that is consistent with imitation
and a fraction $\nu$ are strategy changes not expected from imitation. 
Fig.\ \ref{figspatial} reveals that this approach can capture the average cooperation level in the behavioral experiment. 
Comparing 15 simulations with 15 experimental treatments reveals no significant difference between the simulations and the experiments after Bonferroni correction which takes into account multiple comparison. 
We can summarize this approach by the following equation governing strategy choice, 
\begin{equation}
p_{A \to B} = \nu_{0} \Gamma^{t-1} + \left(1-2 \nu_{0} \Gamma^{t-1} \right) \Theta(\pi_{B} - \pi_{A}),
\end{equation}
where $B$ is the best performing neighbor of $A$, $t$ is the round of the game, $\pi_{A}$ and $\pi_{B}$ are the payoffs of $A$ and $B$, respectively, and  $\Theta(x)$ is the Heaviside function ($\Theta(x)=0$ for $x \leq 0$ and $\Theta(x)=1$ for $x>0$).
In our experiment, we find $\nu_{0} = 0.380 \pm 0.013$ and $\Gamma = 0.962 \pm 0.003$, see Fig. \ref{figtime}.

\begin{figure}[h]
\begin{center}
	\includegraphics[width=\linewidth,angle=0]{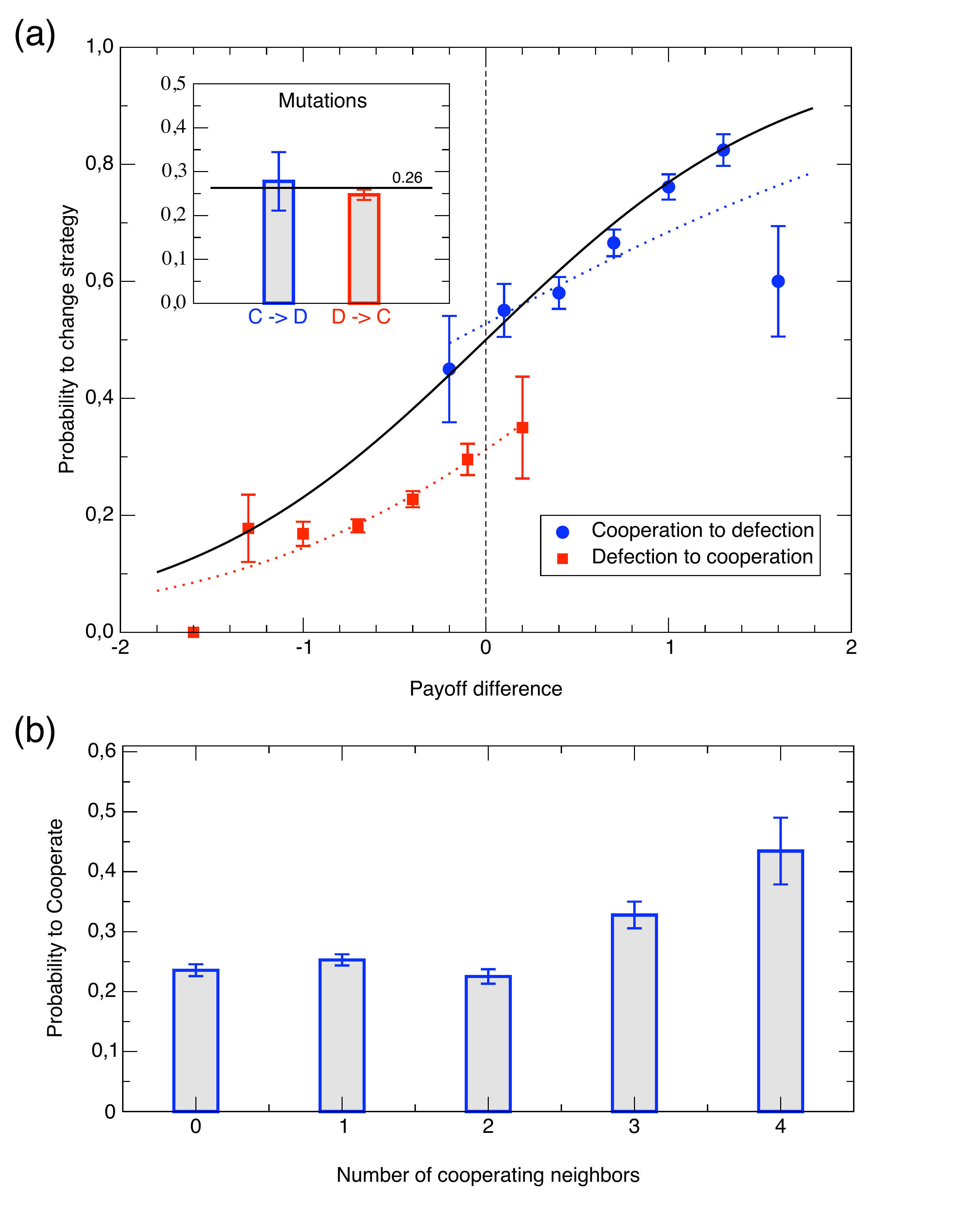}
	\end{center}
\caption{ 
Strategy updating in a spatial game. 
(a) As expected, the probabil0ity to switch to another strategy increases with the payoff difference. 
Theoretical models typically assume strategy update functions such as e.g.\ 
$p= \left(1+\exp \left[-\beta \Delta \pi \right] \right)^{-1} $, where
$p$ is the probability to switch strategy and $\Delta \pi$ is the payoff difference. 
Fitting this function leads to an intensity of selection
$\beta =1.20 \pm 0.25 $ (full line). 
However, the data for cooperating and defecting players seem to follow different characteristics 
and defecting players seem to be more resilient to change than cooperators. 
To capture this, we have also fitted the two different data sets to the function $\left(1+\exp \left[-\beta \Delta \pi + \alpha \right] \right)^{-1} $ (dotted lines).
This approach leads to $\beta_{C} = 0.67 \pm 0.28$ and $\alpha_{C} = -0.11 \pm 0.23$ for cooperating players.
For defecting players, we find $\beta_{D} = 0.99 \pm 0.23$ and $\alpha_{D} = 0.79 \pm 0.14$. 
The inset shows the probability to change strategies spontaneously, without any role
models playing a different strategy. Such spontaneous changes correspond to mutations and occur with probability 
$0.28 \pm 0.07$ (cooperating players switching to defection) or $0.25 \pm 0.01$ (defecting players switching to cooperation). 
This probability is much higher in our experiment than typically assumed for theoretical models, but decreases exponentially in time (see Fig. \ref{figtime}).  
(b) Another perspective is to infer the probability to cooperate in the next round, given the number of coopering neighbors
in the current round. This probability is highest if all neighbors cooperate, although in this case the payoff from defection
would be highest. This indicates that humans do not only imitate what is successful, but also what is common
(All error bars are the standard deviations of a binomial distribution, $\sqrt{p(1-p)/n}$, where $n$ is the number of samples). 
}
\label{figFermi}
\end{figure}

Next, let us abstract from the fact that strategy adoption changes over time and analyze the way in which individuals imitate their co-players in more detail. 
First, we analyze all situations in which players do the same as their four neighbors. 
How likely are they to switch strategies? 
It turns out that cooperators switch to defection in such a homogeneous environment with probability 
$\mu_{C} = 0.28 \pm 0.07$ (averaged over 45 such situations).
Defectors switch to cooperation with probability $\mu_{D} = 0.25 \pm 0.01$ (averaged over 1400 such situations). 
These probabilities correspond to spontaneous mutations or strategy exploration of the players. 
To analyze imitation is less straightforward, because it is impossible to say if people changed to a different
strategy imitating a particular neighbor, several ones at the same time, at random or based on some more sophisticated argumentation. 
For example, human players who find themselves in a neighborhood of cooperators may be
tempted to defect, anticipating to win the highest possible payoff, before another neighbor defects. 
They may also expect others to take advantage of a cooperative neighborhood sooner or later. 
However, we can at least quantify the average behavior. 
We take all decisions into account in which a focal cooperator had at least one defecting neighbor (1524 decisions)
or in which a focal defector has a least one cooperating neighbor (2791 decisions).  
Some of these strategy changes will again correspond to random strategy exploration, but 
we can assume that this occurs with a probability that is independent of the payoff difference. 
 
Depending on the payoff difference to the neighbor who performs best by using a different strategy than the focal player, 
what is the probability that the focal player switches to that other strategy?
Fig.~\ref{figFermi} shows that the probability increases with the success of the neighbor, as expected. 
A cooperator is typically confronted with defector performing better,
while a defector can typically only choose to imitate a cooperator performing worse.
Moreover, defectors are more resilient to change than cooperators. 
 To model strategy changes, we assume that the probability to switch strategy is given by 
$p= \left(1+\exp \left[-\beta \Delta \pi  \right] \right)^{-1} $.
Note that for $\beta \to \infty$, we recover the unconditional imitation from above. 
Fitting this function to the data shown in Fig.~\ref{figFermi} leads to 
$\beta =1.20 \pm 0.25 $. The error corresponds to the standard deviation in a binomial distribution, $\sqrt{p(1-p)/n}$, where $n$ is the number of samples. 
If we want to take the difference in strategy adoption of cooperating players and defecting players into account, 
we can also fit two different functions to the data, see Fig.~\ref{figFermi}
If we instead use the average payoff difference to players using a different strategy, we obtain $\beta = 1.15 \pm 0.23$. 
Also in this case, defecting players seem to be more resilient to change. 

Fig. \ref{figFermi} also shows how the probability to cooperate depends on the number of cooperating neighbors.
This does not take any payoffs into account and addresses wether players imitate the common rather than the 
more successful. It turns out that the probability to cooperate is below $50 \%$ even when all neighbors are cooperating.
Thus, in our experiment players do not only imitate the most common strategy, but decide for cooperation or defection in more complex ways.

The intensity of selection measured in our experiments reveals that humans do not
simply accept any strategy that is performing better than their strategy, as assumed by imitation
dynamics. However, $\beta $ is also so high that analytical results obtained under weak selection
may not always apply.
Again, we can summarize our approach by a simple equation. If neglect temporal dependence, but take the differences between cooperators and defectors into account, we find 
\begin{eqnarray}
p_{C \to D} &=& \mu_{C} + \frac{1-\mu_{C}}{1+e^{-\beta_{C} (\pi_{D} - \pi_{C} ) + \alpha_{C} }} \\
p_{D \to C} &=& \mu_{D} + \frac{1-\mu_{D}}{1+e^{-\beta_{D} (\pi_{C} - \pi_{D} ) + \alpha_{D} }}.
\end{eqnarray}
Our analysis leads to $\mu_{C} = 0.28 \pm 0.07$, $\beta_{C} = 0.67 \pm 0.28$ and $\alpha_{C} = -0.11 \pm 0.23$ for cooperating players
and $\mu_{D}=0.25 \pm 0.01$, $\beta_{D} = 0.99 \pm 0.23$ and $\alpha_{D} = 0.79 \pm 0.14$ for defecting players.

\section{Discussion}

As expected, players imitate others with probability increasing with the payoff difference. 
In evolutionary game dynamics, this corresponds to selection. 
But sometimes players switch spontaneously to a new strategy at random, which corresponds to a mutation.  
Our approach reveals that the probability of such random changes is much higher than typically
assumed in theoretical models.

Theoreticians are often
interested in the dynamics for very large populations and not in finite size effects. However, considering large population is unfeasible in behavioral experiments, where many repeats are required. Moreover, our predecessors lived in small social groups and our behavior may have adapted to that situation. Regardless of the complexity of our modern society, human interactions occur typically within small social groups even today. 
Most importantly, the way players choose strategies based on local information does not seem to be fundamentally different in larger systems \cite{anxo:2009aa}.
Decision making in humans is certainly a complicated process that goes far beyond the simple models that are typically considered. 
However, we argue that important aspects of human behavior are not captured by the different mechanisms of imitation. 
Modeling these processes by random strategy choice can lead to very different dynamics in theoretical models and captures the general trend of the dynamics in our system, cf.\ Fig.\ \ref{figspatial}.

In our experiment, we have analyzed the simplest system in which humans play a spatial game. 
Many challenges lie ahead: 
Theoretical models describe not only interactions on regular lattices, but also heterogeneous networks
\cite{santos:2005pm}, dynamical networks \cite{pacheco:2006pa} or set structured populations \cite{tarnita:2009df}. 
It would be fruitful to initiate a discussion in the scientific community how such more complex models can be approached by behavioral experiments.

\section{Methods}
From 2003-2004, voluntary human subjects for the experiment were recruited from first semester biology courses at the Universities of Kiel, Cologne and Bonn. 
A total of 400 students participated in the experiment. 
The students were divided into 25 groups consisting of 16 players each. 

In the spatial treatment (15 groups) the 16 subjects were virtually arranged on a spatial grid with periodic boundary conditions. 
This torus shaped geometry ensures that there are no edges in the system. 
Each subject had four fixed direct neighbors  throughout the experiment (von-Neumann neighborhood). 
To ensure the players anonymity, each player was identified by a letter ranging from $a$ to $p$ (e.g. $a$ has the following neighbors: $b$, $d$, $e$ and $m$). 
The subjects would exclusively interact with these four neighbors and received no further information about the remaining 11 subjects. 
In the non-spatial control treatment (10 groups) the 16 subjects where positioned on a new random position on the lattice in each round, such that the probability that another interaction with a particular co-player takes place is $4/15$. 
Otherwise the control experiment was conducted exactly in the same way as in the spatial treatment. 
The students were fully aware of whether they were in a fixed or randomized neighborhood. 

The subjects started in both treatments without money on their account. 
Each group played a total of 25 prisonerÕs dilemma rounds, allowing them to earn on average between 10.00 \EUR (for full defection) and 30.00 \EUR (for full cooperation). 
A single player, however, may theoretically also obtain nothing (if the player always cooperates, but his four partners always defect) or up to 40.00 \EUR (if the player always defects, but his partners always cooperate). 

Each subject had a decision box on his/her private table that was equipped with silent YES, NO and OK buttons. During a short oral introduction the subjects received information about the use of their decision box and how their anonymity would be ensured throughout and after the experiment. 
At the beginning of the experiment a written instruction explaining the game (see Supporting Information) was projected on a screen visible to all players. 
Each subject had to confirm via the OK button that he/she had finished reading and had understood each of the displayed instruction pages. 

In both treatments, each subject had to make a single decision in each round Ð either cooperate or defect in the PrisonerÕs Dilemma played with all four neighbors simultaneously. This setting corresponds to synchronous strategy adjustment. 
After every round the subjects could observe the results of the round on their personal display which could display a maximum of 32 characters. Decisions were displayed in the following form:
\begin{equation}
\nonumber
\begin{array}{|c|c|c|c|c|c|c|c|c|c|c|c|c|c|c|c|}
\hline
s &	Y & \hspace{0.15cm} & \hspace{0.15cm}	 &		t  & Y & \hspace{0.15cm} &		u &	N & \hspace{0.15cm}	&	v &	N & \hspace{0.15cm}  &		w &	Y & \hspace{0.15cm}\\	
\hline
  &	6 & &  &		1 & 2 & &		   &	7 &	&	1 &	0 & &		   &	6 & \\
\hline
\end{array}.
\end{equation}
The display has been explained in detail in three examples and subjecs had no problems understanding it.
Here, $s$, $t$, $u$, $v$ and $w$ are the codes for the different players. Each player is provided with 
his own strategy (cooperation, $Y$, or defection, $N$) and payoff as well as the chosen strategies of the direct neighbors and their respective payoffs, which resulted from their interactions with their 4 neighbors (e.g. own payoff of $s$: $6 = 0.60$ \EUR; payoff player $t$: $12 = 1.20 $ \EUR). 
The computer calculated the individual's payoff from all four encounters and transfered the cumulated payoff to the player's account after each round. 
At the end of the experiment the players received the money on their respective account in cash without losing their anonymity,
see \cite{milinski:2005aa} for details. 

Throughout the experiment the complete anonymity of the subjects was assured by the following measures: 
Subjects were seated between separations, such that no visual contact between them was possible.
All boxes were connected to a computer to record each individual decision. The subjects were informed that they were not allowed to talk or to contact each other during the experiment. Each player could only be identified by his pseudonym (a-p) both by other players as well as by the experimenters. 
Pseudonyms could not be connected with the students' real identity.

\section{Acknowledgement}
\small
We are grateful to S.\ Bonhoeffer for helping us to choose appropriate parameters for the experiment. 
We thank T.M.C.\ Bakker, H.\ Arndt, and H.\ Brendelberger for support and the 400 students for their participation, as well as
D. Helbing and A. Sanchez for stimulating discussions.
A.T. is supported by the Emmy-Noether program of the DFG.


\end{document}